\documentclass[twocolumn,tighten]{aastex631}
\usepackage{amsmath}
\usepackage{amssymb}
\usepackage{xspace,xcolor}
\usepackage{natbib}
\usepackage{outlines}
\usepackage{savesym}
\usepackage{booktabs}
\usepackage{longtable}
\usepackage{tikz}
\usepackage{graphicx}
\usepackage{hyperref}
\savesymbol{tablenum}
\usepackage{siunitx}
\usepackage{comment}
\usepackage{float}
\usepackage{braket}

\newcommand{\beq}{\begin{equation}}
\newcommand{\eeq}{\end{equation}}

\newcommand{\JHU}{William H. Miller III Department of Physics \& Astronomy, Johns Hopkins University, 3400 N Charles St, Baltimore, MD 21218, USA}
\newcommand{\ARCO}{Astrophysics Research Center of the Open University (ARCO), Ra’anana 4353701, Israel}
\newcommand{\OUI}{Department of Natural Sciences, The Open University of Israel, Ra’anana 4353701, Israel}
\newcommand{\JHUAM}{Department of Applied Mathematics and Statistics, The Johns Hopkins University, Baltimore, MD, USA, 21218}

\begin{document}

%\maketitle
\title{The Structure of Poloidal Fields Embedded in Thin Disks}
\correspondingauthor{yzenati1@jhu.edu}

\author[0000-0002-0632-8897]{Yossef Zenati}
\affiliation{\JHU}
\affiliation{\OUI}
\affiliation{\ARCO}

\author[0000-0002-2307-3857]{Ethan T. Vishniac}
\affiliation{\JHU}

\author[0000-0003-3370-105X]{Amir Jafari}
\affiliation{\JHUAM}
%\affiliation{\UWD}

%\date{July 2025}

\begin{abstract}
Many accreting systems are modeled as geometrically thin disks. Simulations of accretion disks cannot, at present, be extended to this regime, although local models can address the behavior of narrow annuli within the disk. A global model needs to account for the interactions between a large scale poloidal field, accreted from the environment, and the disk. The disk magnetosphere can be modeled subject to the boundary conditions imposed by the disk. These depend on the structure of the magnetic field as it crosses the disk, and the degree to which the disk can support a sharp bend in the field lines. Building on earlier work we derive a set of equations which describe a stationary disk with an embedded poloidal field. We derive a modified induction equation that incorporates tensorial turbulent diffusivities and a helicity-regulated $\alpha$-effect. We quantify how helicity conservation introduces a nonlinear backreaction on the large-scale dynamo, dynamically coupling turbulent diffusion and $\alpha-$quenching. We discuss the challenges encountered in finding a unique solution under stationary flows $E_\phi =0$, which balances the inflow of $B_z$ due to accretion, the outflow due to radial diffusion of $B_z$, and the vertical movement of $B_r$ due to turbulent diffusion and buoyancy. The vertical profiles of both the azimuthal diffusion coefficient $D_{ijk}$ and the helicity-driven $\alpha_{ij}$ demonstrate that moderate changes in the radial gradient can lead to significant restructuring of the magnetic field geometry. The ability of disks to sustain large bending angles in the poloidal field implies that angular momentum flux through the magnetosphere can dominate over internal transport even for relatively weak fields. Multiple competing factors can result in non-unique solutions, necessitating extra constraints for clarity. The effective use of analytical theory and diagnostic values is crucial, highlighting the role of isotropic turbulence and helicity regulation in magnetized disk environments. 

\end{abstract}

\section{Introduction} \label{sec:intro}

Axisymmetric equilibria provide a useful organizing framework for magnetized accretion flows. In stationary, axisymmetric, ideal MHD the magnetic field lies on surfaces of constant flux function, and the Grad–Shafranov (GS) equation describes the balance of pressure, magnetic, and centrifugal forces; its relativistic counterpart is widely used for black-hole magnetospheres \citep{AbramowiczM+78,HeyvaertsJ_NormanC89,Kluzniak&Kita00,Spruit10,UzdenskyD05,Rodman&Reynolds24}. When a geometrically thin disk threads a large-scale poloidal field accreted from its environment, the exterior magnetosphere can be solved subject to boundary conditions imposed by the disk, including the inclination and shear of the field at the disk surface \citep[e.g.,][]{BlandfordPayne82,PudritzNorman83,Hawley_Krolik01,AndersonJM+05,McKinneyJ_Blandford09}.

Large-scale poloidal magnetic flux is a controlling ingredient in accretion-powered outflows and jet production, yet its transport and geometric embedding in geometrically thin disks remain poorly constrained. Global simulations can not yet access the extreme thin-disk regime, while local shearing-box models do not determine how externally supplied vertical flux is redistributed over many radii \footnote{The context of magnetic surfaces and the magnetic flux function are crucial components for understanding the behavior of plasma and magnetic fields in astrophysical sources with complex magnetic field geometries \citep{Spruit13}.}. 
A practical global theory, therefore, needs a closure for how MRI turbulence advects, diffuses, and reorients an imposed poloidal field. It must provide disk-imposed boundary conditions for the exterior, approximately force-free magnetosphere. Near black holes, the plasma is typically highly ionized and therefore electrically conducting, leading to strong coupling between the disk and the embedded magnetic field \citep{BlandfordZnajek77, ContopoulosI+99,BelenkayaE2012, PodgornyJ+23,VyasPeer25}. In the inner accretion flow, Spitzer conductivities imply $\mathrm{Rm}\gg 1$, so ideal MHD is appropriate for the bulk dynamics \citep{BalbusHawley98, Frank_King_Raine02}. However, resistive layers and reconnection can be dynamically relevant \citep{QianQ17, NathanailA+25}. The global transport of vertical flux in thin disks is set by the competition between inward advection and turbulent diffusion, both of which are anisotropic under MRI turbulence. The work of \citet{LPP94} emphasized vertical mixing of radial and azimuthal magnetic field components across the midplane, leading to rapid outward diffusion of the poloidal field and limiting bending angles to values of order the disk thickness divided by its radius, $H/r$. Taken at face value this implies very little central concentration of the flux. Later work showed that vertical structure, surface layers, and anisotropic transport can favor inward flux accumulation \citep{Spruit10,RothsteinD+08,GuiletJ_OglivieG13,LiJ_CaoX19,LiJ_CaoX25}. However, \citet{JV18} demonstrated that magnetic buoyancy can lead to strong field line bending in a thin accretion disk even in the presence of nearly isotropic turbulence and without considering the effect of surface layers. Finally, direct simulations of disks further reveal anisotropic turbulent electromotive forces that concentrate $B_z$ into zonal structures, effectively anti-diffusing vertical flux in some regimes \citep{FromangStone09,LizanoS+16}. 
The central difficulty is that thin disks can satisfy the stationary constraint $E_\phi=0$ in multiple ways. In the usual kinematic picture, inward advection of $B_z$ by accretion competes with outward turbulent diffusion of $B_z$, while vertical transport of the inclined field, and magnetic buoyancy, further modifies the balance. Since these processes can counterbalance each other, stationary solutions are not necessarily unique. The induction constraint can be met by a range of flux profiles, from cases with minimal field-line bending, where magnetic buoyancy is insignificant, to the maximum estimates of
%Because these processes can offset each other, stationary solutions need not be unique, the induction constraint can be satisfied by a continuum of flux profiles, from limiting cases with minimal bending of field lines where magnetic buoyancy is negligible to the maximal estimates of 
\cite{Jafari_Vishniac_Vaikundaraman19} in which radial diffusion is ignored. This degeneracy implies that the poloidal-field geometry and in particular the surface inclination relevant to launching winds \citep{BlandfordPayne82,PudritzNorman83,AndersonJM+05,McKinneyJ_Blandford09} can not be determined from advection--diffusion arguments alone without additional physical constraints. This indeterminacy can be resolved by imposing a vertical boundary condition on the poloidal field, but that in turn requires a detailed model for the vertical structure of the large scale poloidal field within the accretion disk.

In this work, we derive a stationary thin-disk model in which the turbulent electromotive force is closed using a tensorial turbulent diffusivity $D_{ijk}$ and a helicity-regulated dynamo tensor $\alpha_{ij}$. Magnetic helicity conservation at high conductivity couples the $\alpha$-effect to turbulent diffusion, introducing a nonlinear backreaction that self-regulates the mean-field response. In the presence of MRI-driven anisotropy, the resulting transport coefficients vary strongly with height. They can reverse the effective radial transport of vertical flux, enabling flux concentration rather than dispersion. We then impose a physically motivated vertical matching condition at the disk surface, $P_{\rm mag}=P_{\rm gas}$, to connect the disk solution to an exterior magnetosphere and determine the resulting field-line bending.

These results bear directly on whether thin disks can accumulate the large-scale flux required for magnetically arrested states and powerful jets \citep[e.g.,][]{TchekhovskoyA+11, WhiteC+19a}, and more generally on how state-dependent poloidal-field geometry may arise in accreting black-hole systems.

This paper is organized as follows. We introduce and provide the basic argument and an overview of the problem and the physical motivation in Section \ref{sec:overview}. The setup of our theoretical model and equations focus on their effects in Section \ref{sec:model}. In Section \ref{sec:Result}, we present the solutions and their dependence on the model parameters, along with a discussion of uncertainties and limitations. Additionally, we focus on the latter case and the resulting properties of the vertical disk structure. In Section \ref{sec:obesr} we consider observational implications. In Section \ref{sec:conclusion}, we summarize our conclusions.

\begin{deluxetable}{ll}
\tablecaption{Relevant turbulent transport quantities used in the MHD closure model.}
\tablehead{
\colhead{Symbol} & \colhead{Description}
}
\startdata
$\alpha_{ij}$ & Helicity-regulated turbulent dynamo coefficient \\
$\bar{Q}_{ij}$ & Components of the turbulent anisotropy tensor \\
$\epsilon_{ijk}$ & Levi-Civita symbol (antisymmetric tensor) \\
$D_{ij}$ & Turbulent Reynolds stress tensor \\
$M_{ij}$ & Turbulent Maxwell stress tensor \\
$\partial_t H$ & Temporal evolution of magnetic helicity \\
$D_\phi$ & Azimuthal component of turbulent diffusion \\
$\lambda_{\eta}$ & Kolmogorov microscale $\lambda_{\eta} \sim L R_e^{-3/4}$ \\
\enddata
\end{deluxetable}

\section{Basic Arguments}
\label{sec:overview}

The model developed in this work addresses the magnetic structure of thin, turbulent accretion disks under stationary, axisymmetric conditions. Its core formulation arises from the modified induction equation, where both the turbulent electromotive force and dynamical backreaction are resolved using tensorial structures. The physical system is governed not only by mean-field MHD but by the requirement that magnetic helicity is conserved in high-conductivity flows. This constraint introduces a nontrivial coupling between the turbulent diffusivity tensor $D_{ijk}$ and the helicity-regulated dynamo tensor $\alpha_{ij}$, rendering the field evolution intrinsically nonlinear.

Rather than assuming a prescribed $\alpha-$effect or isotropic turbulent diffusion, the model derives both quantities from the response of anisotropic magnetorotational instability (MRI)-driven turbulence to the evolving mean field. Likewise, the anisotropic turbulence composed of eddies that are azimuthally elongated and vertically compressed, leading to a turbulent eddy wavenumber anisotropy tensor $\bar{Q}_{ij}\equiv \overline{\hat{k}_i\,\hat{k}_j}$ %$\bar{Q}_{ij}\equiv \bar{\hat{k_i}\hat{k_j}}$
characterized by $\bar{Q}_{\phi\phi} \ll \bar{Q}_{zz}, \bar{Q}_{rr}$. This structure critically modifies the effective turbulent transport of magnetic flux. The anisotropy, encoded in the tensor $\bar{Q}_{ij}$, reflects the geometry of elongated azimuthal eddies and suppressed vertical motion. In this setting, the vertical magnetic field component $B_z(r,z)$ becomes dynamically regulated by the competition between inward advection from accretion, outward turbulent diffusion, and vertical transport due to buoyancy and turbulent pumping. The key physical mechanism is that helicity conservation forbids unbalanced transfer between large and small-scale fields. 
In steady state, this enforces a vanishing net helicity flux, resulting in a dynamically determined $\alpha_{ij}$ that reflects the background field geometry and turbulent anisotropy equation \ref{eq:alphaij_diff}. Consequently, the turbulent diffusion and field amplification processes are no longer independent; they self-regulate to maintain helicity balance.

This framework leads to qualitatively new behavior. In regions near the disk midplane, where turbulence is most substantial and anisotropy most extreme, the radial diffusion of vertical magnetic flux can reverse sign, leading to flux concentration rather than dispersion \citep{BaiX_StoneJ14,LiJ_CaoX19,ReaD+24}. The model thus predicts localized magnetic structures without invoking internal dynamo growth or external boundary conditions. Instead, the magnetic field configuration emerges from a self-consistent balance between turbulent transport, helicity regulation \citep[e.g,][]{Vishniac_Cho01,Blackman04}, and the global constraints of the GS equilibrium. Unlike classical $\alpha-$ disk models or kinematic dynamos, this formulation does not assume field growth; instead, it predicts steady magnetic field profiles arising from constrained transport. The vertical magnetic structure is therefore not a free parameter but an emergent feature determined by the geometry of anisotropic turbulence, the balance of inflow and diffusion, and the suppression of helicity exchange. This provides a new route toward understanding disk-magnetosphere coupling in systems where turbulence coexists with coherent vertical flux.

\section{The Model} \label{sec:model}

In this section we will describe our formalism, derive the governing equations and examine the effects of turbulence. We start with the equations of ideal MHD. Dissipative effects will be mediated only by turbulence. We will use the assumption of an axisymmetric thin disk to reduce the full equations to a useful model. First we note that the ratio of the disk height to its radius is roughly
\begin{equation}
{H\over r}\sim {c_s\over r\Omega(r)}\ll 1,
\end{equation}

where $H$ is the disk thickness, $c_s$ is the sound speed, and $\Omega(r)$ is the angular velocity. In general we can assume that derivatives with respect to radius can be ignored, with exceptions that will be noted.

Second, at any radius we can scale the vertical component of the magnetic field with the disk midplane pressure, $B_z / \sqrt{P}|_{(z=0)}$. We use this in place of the usual definition of magnetization, i.e. $B_z/\int\rho dz$, since it is the dynamically relevant local quantity. We also set the boundaries of disk at $P_{mag}=P_{gas}$.
When the local magnetization is of order unity the magnetosphere has, by definition, swallowed the disk altogether. 

Third, we will ignore the vertical motions of the plasma in the disk. In other words, we ignore the disk wind and assume that the plasma is always in vertical equilibrium. An estimate of the outflow will be left for future work. However, a radial flow $v_{acc}$ is essential as long as $v_z$ is of order $(H/r) v_r$ or smaller. Since $v_z$ vanishes by symmetry at the midplane this implies that $v_z$ is dynamically unimportant throughout the thin disk. It becomes important only when gas pressure is negligible and the magnetic forces are balanced only by inertial effects \citep{Hawley&Balbus02,Kato+book08,SorathiaK+12,LizanoS+16,Jacquemin-Ide+21}.

% Having decided to ignore the wind the following paragraph can be deleted
%Conservation of mass requires
%\begin{equation}
%    \nabla\cdot(\rho{\bf v})=0.
%\end{equation}
%Assuming axisymmetry, we only need to consider radial and vertical motion. For radial velocities comparable to vertical motions, we have 
%\begin{equation}
%   \rho v_z = constant.
%\end{equation}

%With a typical system, we can derive from the induction equation the magnetic Reynolds number $Re_{m} = 4\pi\sigma_0 L \langle v \rangle/c^2 > 1$. 
%However, a radial flow $v_{acc}$ is essential as long as $v_z$ is of order $(H/r) v_r$ or more minor. Since $v_z$ vanishes by symmetry at the midplane, this tells us that $\rho v_z$ only tends to a constant away from the midplane \citep{Hawley&Balbus02,Kato+book08,SorathiaK+12,LizanoS+16,Jacquemin-Ide+21}.

\subsection{Equilibrium conditions}

For any magnetic field {\bf B} we have
\begin{equation}
    \nabla\cdot{\bf B}=0,
\end{equation}
so $\partial_zB_z\approx0$.
Within the thin disk, we can take $B_z$ equal to a function $B_z(r)$.

%Notably, looking at the motion equation $\rho \left(\partial_t v + \bf v \cdot \nabla \right)= -{\bf \nabla} p + \frac{\pi}{4}\left(\bf {\nabla \times B}\right)\times {\bf B} +\rho \bf g$ and neglecting radial gradients, 

Neglecting radial gradients, the radial force equation is

\begin{equation}
    v_z\partial_z v_r=r(\Omega^2-\Omega_{kep}^2)
    +\frac{1}{\mu_0}{B_z\partial_z B_r\over\rho},
\end{equation}
where $r\Omega_{kep}^2$ is the inward gravitational acceleration.
Ignoring the inertial term this becomes

\begin{equation}
\Omega^2=\Omega_{kep}^2-\frac{1}{\mu_0}{B_z\partial_zB_r\over\rho r},
\end{equation}

This implies the possibility of significant vertical shear stretching the vertical field lines.

\begin{equation}
B_z r\partial_z \Omega=-{B_z^2\over 2\mu_0\Omega}\partial_z\left({1\over\rho}\partial_z B_r\right),
\end{equation}

This will be more important when the vertical scale height for $B_r$ is much less than the density scale height, in which case we can write

\begin{equation}
B_z r\partial_z \Omega=-{B_z^2\over 3\mu_0P}
z|L_p|{\partial_z^2B_r\over B_r}({3\over2}B_r\Omega),
\end{equation}

where the last factor is the rate of generation of $B_\phi$ from the shearing of the radial field.  We see that 
this can have a large effect on the evolution of $B_\phi$ even if $\Omega$ is always close to $\Omega_{kep}$, but only for magnetic scale heights small compared to the disk thickness and significant magnetization.

Similarly, the vertical force equation is

\begin{equation}
    {1\over2}v_z^2+z\Omega_{kep}^2+{1\over\rho}
    \partial_z\left(P+\frac{B_r^2+B_\phi^2}{2\mu_0}\right)=0,
\end{equation}
Ignoring vertical plasma motions this just becomes the condition for vertical hydrostatic equilibrium.  In this paper we will ignore the magnetic pressure term since we truncate the disk when the magnetic pressure is equal to the gas pressure.

In the same limit the azimuthal force equation, which is also the angular momentum conservation equation, is

\begin{equation}
    {1\over r}\partial_r( F_{MRI})+{v_r\rho\over 2}\Omega=\frac{1}{\mu_0}\,B_z\partial_z B_\phi,
    \label{eq:amflux}
\end{equation}

where $F_{MRI}$ is the radial angular momentum flux density due to MRI turbulence, here $\Omega >0$ \citep{GellertM+12,RudigerG+15,HeldL+24}. The radial derivative here remains significant even in the thin disk limit. The amplitude of the MRI transport term has a minimum value due to self-sustained turbulence driven by the MRI, but is proportional to the vertical field for sufficiently large $B_z$.

To solve the structure of the magnetic field inside the disk, we need to solve the induction equation for a stationary system,

\begin{equation}
    \partial_t \bf {B} = \nabla\times \bf {V}\times \bf{B} + \eta_m \nabla^2 \bf{B} = -\bf{B} \nabla \cdot v + \left(\bf{B} \nabla \right) \mathbf{v} = 0, \label{eq:Induc}
\end{equation}

Here, the velocity $\bf V$ is not just the plasma velocity but also the separate drift of the magnetic field through the plasma due to the effects of turbulence, and magnetic diffusivity is $\eta_m \equiv 1/\mu_0 \kappa_e$.

In general, $\partial_t{\bf B}=0$ implies $\nabla\times{\bf E} = 0$ as a result; that could describe how the ratio between magnetic flux and the mass density changes following the fluid velocity uneven along a field line (Walen's equation) \citep{LizanoS+16,ChanHo+18}. However, if the total $B_z$ flux within any radius is constant, then $E_\phi =0$, whose implications for the radial transport of vertical fields were discussed in \cite{JV18}. This, plus the condition of azimuthal symmetry, means that the condition that $B_r$ and $B_z$ are stationary is automatically satisfied \citep{PiotrovichM+16,Jacquemin-Ide+2019,Sharda+21}. Consequently, the induction equation gives us just one condition

\begin{equation}
    \partial_tB_\phi=-\partial_z E_r+\partial_r E_z-{E_z\over r}=0 ; \label{eq:induction_res}
\end{equation}

Within a thin disk, in a locally co-rotating frame of reference, this becomes

\begin{equation}
    0=r\left(B_r\partial_r\Omega+B_z\partial_z\Omega\right)-\partial_z E_r,
\end{equation}

\subsection{The effects of turbulence}

In the presence of turbulence, the electric field is usually written as

\begin{equation}
    -{\bf E}=\epsilon_{ijk}V_jB_k+\alpha_{ij}B_j+D_{ijk}\partial_jB_k,
\end{equation}

where we use implicit summation. We assume that we can separate the dynamics of the disk into the MRI driven local dynamo and the dynamics of the entrained vertical flux. That way, we only include contributions to $\alpha_{ij}$ driven by the poloidal field.

A combination of effects drives the velocity field $V_j$. There will be a vertical component due to the impact of magnetic buoyancy and turbulent pumping \citep{JV18}, which is distinct from the plasma bulk motion. The radial component is due to the accretion of the plasma. Finally, in general there will be $z$ dependent plasma motion will be induced by magnetic forces, leading to a wind \citep{ChanHo+18}, which we will not address here. 

There is a standard procedure for evaluating $\alpha_{ij}$ and $D_{ijk}$, which involves taking the terms in the induction equation and the force equation, which depend on ${\bf B}$ and its spatial derivative and substituting them into $\partial_t\langle {\bf v\times b}\rangle\tau_c$ \citep{Vishniac95,Hawley&Balbus02,Takeuchi&Okuzumi14,HeldL+24}. For $\alpha_{ij}$ we have the standard expression

\begin{equation}
    \alpha_{ij}=\epsilon_{inm}\langle \tilde v_n
    \partial_j\tilde v_m-\rho^{-1}b_n\partial_j b_m\rangle\tau_c.
    \label{alpha}
\end{equation}

Here, $\bf\tilde v$ and $\bf b$ are the turbulent velocities and magnetic fields, and the brackets denote averaging over turbulent eddy scales. The turbulence induced by MRI may lead to both these terms being nonzero. Nevertheless, we will disregard that effect here, not because it is nonexistent but because, for the sake of simplicity, we presume that the dynamo effects induced by the MRI do not interact with the advected poloidal field. We shall incorporate a magnetic contribution to $\alpha_{ij}$, induced by the dynamics of the poloidal field. Similarly, we can calculate $D_{ijk}$. We get,

\begin{equation}
    D_{ijk}\partial_jB_k=-\epsilon_{ijk}\left(R_{jm}+\rho^{-1}M_{jm}\right)\tau_c\partial_m B_k
    \label{eq:dsim}
\end{equation}

where $R_{ij}$ and $M_{ij}$ are the turbulent Reynolds and Maxwell tensors, respectively. In our problem, we expect the $r\phi$ components of these tensors to be nonzero. This expression neglects the pressure term in the force equation, which can have important consequences. For nearly incompressible motions, we can use the condition that $\partial_t\nabla\cdot\tilde v=0$ to write the pressure term as

\begin{equation}
    P={\partial_i\partial_j\over\nabla^2}(-\tilde v_i\tilde v_j+\rho^{-1}(B_i+b_i)(B_j+b_j)).
\end{equation}

The inverse of the Laplacian is not a uniquely defined operator, but if we assume that any boundary contributions are small, it is equivalent to dividing by $-k^2$ in Fourier space.
The operator $\partial_i\partial_j/\nabla^2$ is the product of two Riesz transforms with a negative sign. Alternatively, it's equivalent to multiplying by two unit vectors in Fourier space. For simplicity, we will denote it as $Q_{ij}$ and its expectation value as $\bar Q_{ij}$. In the presence of a large-scale field ${\bf B}$, we can write its contribution to the pressure gradient as 

\begin{equation}
    \begin{split}
        \partial_iP_B={\partial_i\over\nabla^2}2(\partial_mB_j\partial_j b_m)\approx 2\partial_mB_j{\partial_i\partial_j\over\nabla^2}b_m\\
        = 2Q_{ij}b_m\partial_mB_j.
    \end{split}; \label{eq:Pgrad}
\end{equation}

The expression for $D_{ijk}$ in equation(\ref{eq:dsim}) is modified to

\begin{equation}
    \begin{split}
        D_{ijk}\partial_jB_k = -\epsilon_{ijk}\left(R_{jn}\delta_{km} + \rho^{-1}M_{jn}(\delta_{mk}-2\bar Q_{mk})\right)\\
        \tau_c\partial_n B_m ; \label{eq:diffusion}
    \end{split} 
\end{equation}

As a consequence of this change, sufficiently anisotropic eddies can give rise to a negative diffusion term in some directions. We have turbulent eddies elongated in the azimuthal direction for the MRI, so $\bar Q_{\phi\phi}$ is much smaller than other components \citep{MurphyG+15,Hogg&Reynolds18,MeduriD+23}. The relative extent of eddies in the vertical and radial directions can vary depending on the importance of buoyancy and the strength of the perpendicular magnetic field. As long as the azimuthal field dominates, the eddies will range from "flat," i.e., $\bar Q_{zz}\approx 1$, to $\bar Q_{rr}\sim \bar Q_{zz}\sim 1/2$ \citep{MurphyG+15,Jacquemin-Ide+2019}. Consequently, the magnetic diffusion term for radial diffusion of the vertical field will vary from being effectively an "anti-diffusion" term to roughly zero. The difference between equation(\ref{eq:dsim}) and equation(\ref{eq:diffusion}) is never negligible \cite{Jacquemin-Ide+21}.

We get expressions for turbulent diffusion by applying equation(\ref{eq:diffusion}) to an azimuthally symmetric thin disk. First, the azimuthal component is

\begin{equation}
\begin{split}
    D_{\phi jk}\partial_jB_k=-(R_{zz}+\rho^{-1}M_{zz}(1-2\bar Q_{rr}))\\
    \tau_c\partial_z B_r+2\rho^{-1}M_{zz}\bar Q_{r\phi}\tau_c\partial_zB_\phi\\
    +(R_{rr}+\rho^{-1}M_{rr}(1-2\bar Q_{zz}))\tau_c\partial_rB_z,
    \end{split}
    \label{eq:Diff_Azm}
\end{equation}

Where we have assumed that the only non-diagonal terms in any of the tensors are the $\hat\phi\hat r$ terms, the last term in this expression includes a radial derivative, which should be small in a thin disk. However, we cannot discard it at this point because, near the midplane of the disk, we expect that both $B_r$ and $B_\phi$ will be smaller than $B_z$ by a factor of order the height of the disk divided by its radius. Consequently, all the terms in this expression will be comparable near the midplane and to the radial accretion velocity times $B_z$. Achieving a balance between these terms is the key to an accretion disk's stationary distribution of $B_z$. We need to keep them all.

One last modification to the diffusion term arises from the robust conservation of magnetic helicity, ${\bf A}\cdot{\bf B}$, at high magnetic Reynolds numbers. This conservation law implies that the diffusion term will induce a contribution to  $\alpha_{ij}$, as we explain below. While the total magnetic helicity is necessarily conserved, it can be embedded in large-scale or turbulent fields. If we denote the former as $H$ and the latter as $h$, then the evolution of $H$ can be written as 

\begin{equation}
    \partial_t H=2{\bf B}\cdot\langle {\bf \tilde v\times b}\rangle-\nabla\cdot{\bf J_H},
\end{equation}

where $J_H$ is the magnetic helicity flux, that term is an awkward way to rewrite the usual evolution equations for ${\bf B}$ and ${\bf A}$. However, the first term on the RHS side of this equation describes the transfer of magnetic helicity between scales, in this case, between turbulent eddies and the large-scale structure of the field. In a stationary state, this term should be zero. The eddies will accumulate magnetic helicity, force a back reaction on the large-scale field, and drive the magnetic helicity transfer to zero if it is nonzero. From equation(\ref{eq:Alpha_ij}) we see that magnetic contribution to $\alpha_{ij}$ is

\begin{equation}
    \begin{split}
    \alpha_{ij}=-\rho^{-1} \tau_c\epsilon_{inm}\epsilon_{mlq}\langle b_n\partial_j\partial_l a_q\rangle =\\
    \rho^{-1}\tau_c\langle b_n\partial_i\partial_j a_n\rangle\approx \rho^{-1}k^2\tau_c\bar Q_{ij}h,
    \end{split}; \label{eq:Alpha_ij}
\end{equation}

Where $k^2$ is the mean square wavenumber of the eddies. The condition that the transfer of helicity between scales vanishes is
\begin{equation}
    0=B_i\alpha_{ij}B_j + B_iD_{ijk}\partial_jB_k
\end{equation}
Consequently turbulent diffusion induces an $\alpha_{ij}$ term equal to 
\begin{equation}
    \alpha_{ij}=-\bar Q_{ij} \left[{B_qD_{qsu}\partial_sB_u\over B_n\bar Q_{nm}B_m}\right]; \label{eq:alphaij_diff}
\end{equation}

All the turbulent effects will be negligible in regions where the MRI is weak, assuming no other instabilities are present. In terms of constructing a thin disk model, this means that the turbulent contributions to $E_\phi$ have to go to zero at the boundary of the turbulent region for $E_\phi$ to remain zero throughout the disk.

As mentioned above with the induction equation(\ref{eq:induction_res}), the interest regime lies between the inflow of $B_z$ during the accretion and the outflow due to radial turbulent diffusion. In characteristic scale $L$ the steady-state solution is $\partial B_z/\partial t = 0$, and the MRI growth rate is $\lambda \approx 0.5\frac {-d\ln{\Omega}}{d\ln{r}}\Omega$, specifically driven turbulent diffusivity lead to radial scale $L_r \approx \alpha_{\rm MRI} \frac{H^2 \Omega}{v_r}$ and vertical scale $ L_z \approx \alpha_{\rm MRI} \frac{H^2 \Omega}{v_{\rm buoy}}$. Eventually, the characteristic timescale can be driven with the two effective scales and equations \ref{eq:Alpha_ij} and \ref{eq:alphaij_diff} to calculate the $\bar Q_{ij}$ of the $\hat\phi\hat r$ terms.

\subsection{The Vertical Boundary}  \label{vertdisk}

The internal dynamics of a thin disk are contained within range of $z$ which is much less than its radius. Neglecting the effect of the disk wind, the exterior is force-free, i.e. ${\bf J}\times{\bf B}=0$, or equivalently, the GS equation without a pressure term. This equation can be solved for any ${\bf B}$ at the disk boundary, but not without discontinuities in the current. Without a detailed solution for the plasma flow at the surface, and its affects on the magnetosphere, we require a physically plausible boundary condition for the magnetic field.  

We propose to set the boundary surfaces of the disk at $P_{mag}=P_{gas}$ and to ignore vertical equilibrium, requiring only that ${\bf J}\times {\bf B}$ in the horizontal plane be parallel to the component of ${\bf B}$ in that same plane. Ignoring radial gradients, this is the condition that

\begin{equation}
{B_z\partial_z B_r\over B_z\partial_z B_\phi}={B_r\over B_\phi},
\end{equation}
or

\begin{equation}
\partial_z\left(B_r/B_\phi\right)=0.
\end{equation}

In other words, we require that the vertical shear of the horizontal magnetic field vanish at the disk boundary. By neglecting vertical equilibrium we are allowing for the possibility that the boundary layer is being expelled into the magnetosphere. The most significant approximation here is that by assuming the alignment of the horizontal force with the field lines, we are neglecting the inertia of the plasma in the disk wind.

There is a trivial solution with this boundary condition.  For any radial distribution of vertical flux we can assume $B_r=B_\phi=0$ everywhere. We can satisfy all the structure equations by balancing the accretion flow with the outward radial diffusion of $B_z$.  While this is a formal solution to the model it is not physically self-consistent.  In this limit our assumption that the radial gradient in the magnetic pressure is small compared to the radial component of the magnetic tension is violated and this solution can be discarded.

\subsection{$B_z(R)$ Plug In a Isothermal Disk}

From equations \ref{eq:diffusion} and \ref{eq:Pgrad} the pressure included in the $Q_{ij}$ and modified via the diffusion tensor. For Gaussian vertical profile $\rho(z) = \rho_0 e^{\frac{-z^2}{2H^2}}$ and $H/R \ll 1$ , the $B_r,\ B_\phi \sim \mathcal{O}(H/R)B_z$ that led to $\partial_z\sim 1/H$ and $\partial_R \sim 1/R$ are comparable near the midplane. It is straightforward to calculate the properties of the disk including the midplane temperature, surface density etc. All this can be used as an initial guess for $B_z(R)$ before enforcing the helicity-regulated constraint as in equation(\ref{eq:alphaij_diff}). Consequently, the factors $(1-2\bar Q_{zz})$ in equation(\ref{eq:Diff_Azm}) decide whether you get diffusion or flux concentration.

\begin{figure*}
    \centering
   \includegraphics[width=0.49\linewidth]{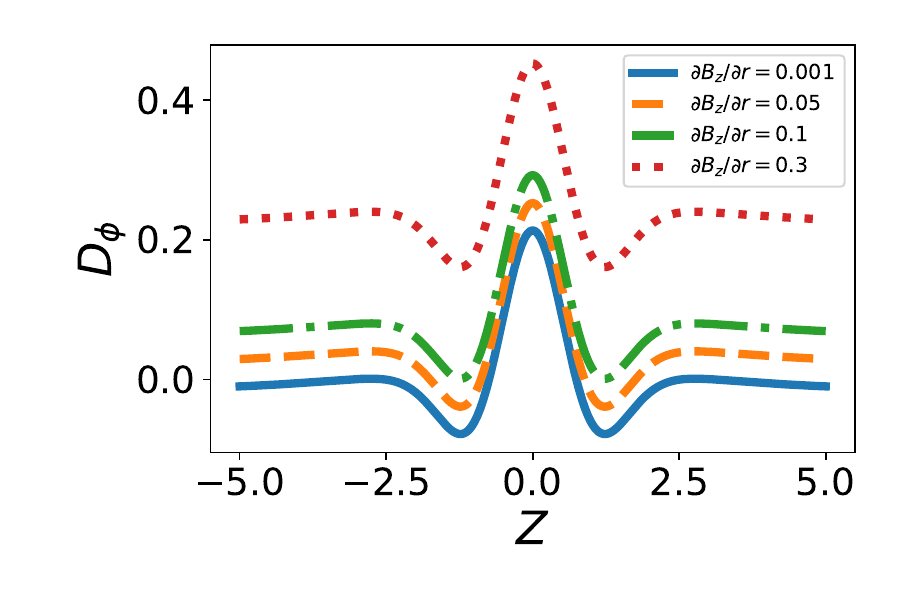}
   \includegraphics[width=0.49\linewidth]{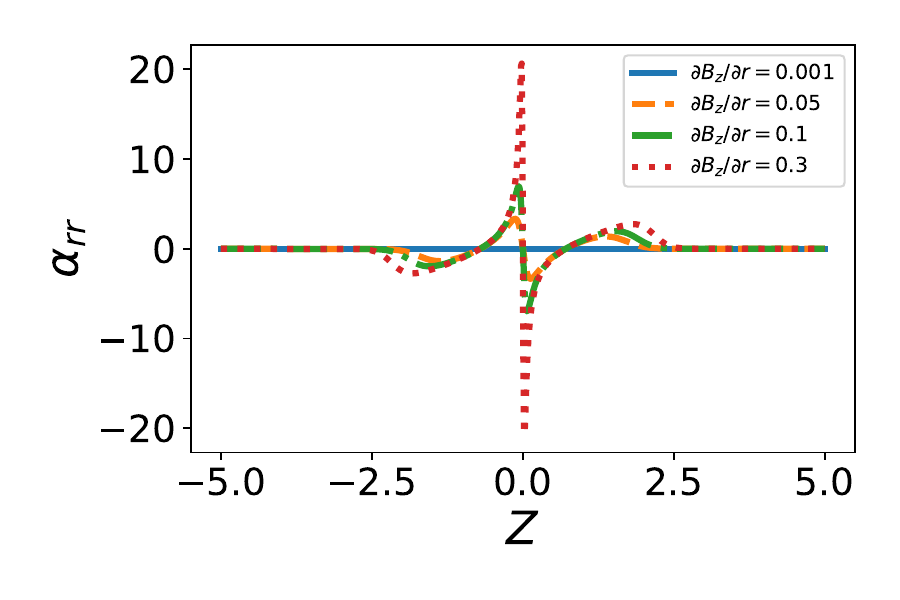}
    \caption{Left: Vertical profile of the azimuthal turbulent diffusion term $D_\phi(z)$ from Equation \ref{eq:Diff_Azm}, plotted for three values of the radial gradient of the vertical field: $\partial B_z/\partial{r} = 0.01$, $0.1$, and $0.3$. The diffusion term includes contributions from vertical derivatives of $B_r$ and $B_\phi$, and a radial derivative of $B_z$. As $\partial B_z/\partial{r}$ increases, the contribution from radial magnetic structure becomes more prominent, especially near the midplane. The magnetic field profile in the vertical direction is Gaussian, and in the azimuthal direction, it is sinusoidal. $B_r$ is derived from the gradient of $B_z$. All the turbulence parameters, Reynolds, Maxwell stress tensor, and anisotropy tensor, taken from direct and fitting simulation data \citep{Jacquemin-Ide+21}. Right: Vertical profile of the helicity-regulated $\alpha_{rr}(z)$ coefficient for three values of the radial gradient of the vertical field, $\partial B_z/\partial{r} = 0.01$, $0.1$, and $0.3$. In this model, the poloidal component $B_r(z)$ is assumed to scale with both $\partial B_z/\partial{r}$ and $\partial_z B_z$, reflecting the coupling of radial and vertical magnetic structures in the induction equation. Larger $\partial B_z/\partial{r}$ values result in stronger poloidal fields, enhancing the helicity-driven backreaction and modulating the $\alpha$-effect accordingly.}.
    \label{fig:Dphi_eq16_alpha_rr_dBzdr}
\end{figure*}

%\begin{figure*}
    %\centering
   %\includegraphics[width=1\linewidth]{alpha_rr_plot_vs_dBzdr.pdf}
    %\caption{Vertical profile of the helicity-regulated $\alpha_{rr}(z)$ coefficient for three values of the radial gradient of the vertical field, $\partial B_z/\partial{r} = 0.01$, $0.1$, and $0.3$. In this model, the poloidal component $B_r(z)$ is assumed to scale with both $\partial B_z/\partial{r}$ and $\partial_z B_z$, reflecting the coupling of radial and vertical magnetic structures in the induction equation. Larger $\partial B_z/\partial{r}$ values result in stronger poloidal fields, enhancing the helicity-driven backreaction and modulating the $\alpha$-effect accordingly.}
    %\label{fig:alpha_rr_dBzdr}
%\end{figure*}

\section{Results \& DISCUSSION} \label{sec:Result}

\subsection{Potential model and overview results}

This phenomenon also enhances the back reaction, resulting in more pronounced saturation effects in regions with steep field gradients. This non-linear behavior highlights the delicate interplay between substantial field growth and helicity-conserving suppression in realistic disk environments. Together, these findings provide quantitative evidence of how anisotropy, field geometry, and helicity conservation collaborate to regulate the magnetic structure of accretion disks.

Figure~\ref{fig:Dphi_eq16_alpha_rr_dBzdr} shows how $D_\phi(z)$ responds to changes in the radial gradient of the vertical magnetic field $dB_z/dr$. Increasing this gradient enhances the outward diffusive transport of vertical flux due to the third term in Equation \ref{eq:Diff_Azm}. The shape of $D_\phi(z)$ remains approximately symmetric about the midplane. Still, it increases in magnitude, indicating that even moderate changes in the radial structure of $B_z$ can lead to significant changes in vertical flux transport. This results from the increasing dominance of azimuthally elongated eddies, which are less efficient at regenerating large-scale poloidal fields. Notably, the profiles become increasingly shallow and quenched near the midplane, demonstrating how MRI-driven turbulence can self-limit field amplification through helicity feedback.

In figure~\ref{fig:Dphi_eq16_alpha_rr_dBzdr} explores how $\alpha_{rr}(z)$ varies with $\partial B_z/\partial{r}$ when the poloidal component $B_r(z)$ is modeled as proportional to both $\partial B_z/\partial{r}$ and $\partial_z B_z$, consistent with expectations from the induction equation. The results indicate that stronger radial gradients in $B_z$ indirectly amplify $B_r$, thereby increasing the strength of the helicity-regulated $\alpha$-term.

Our formulation also complements the work of \citet{JV18}, who analyzed radial flux transport without considering helicity constraints. By including the helicity feedback explicitly in equation \ref{eq:alphaij_diff}, we demonstrate that turbulent transport and $\alpha$-quenching are not independent effects but are dynamically linked.

Further, the anisotropic structure of MRI turbulence reported in simulations by \citet{MurphyG+15,Jacquemin-Ide+21} provides observational and numerical support for our assumptions regarding $\bar{Q}_{ij}$. In particular, our use of an anisotropic $\bar{Q}_{ij}$ to explain field concentration is a natural extension of their findings.

\begin{figure*}
    \centering
   \includegraphics[width=1\linewidth]{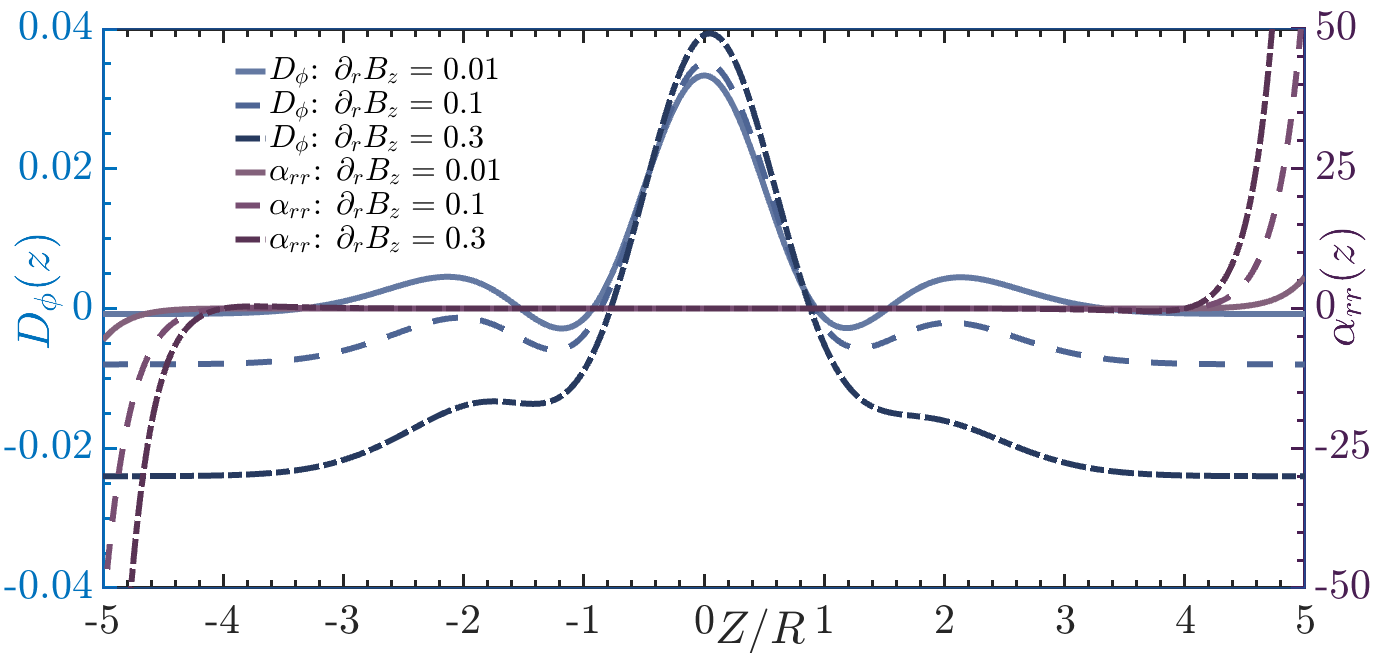}
    \caption{Vertical profiles of the azimuthal turbulent diffusion coefficient, $D_\phi(z)$ (Blue solid, left axis), and the helicity dynamo coefficient, $\alpha_{rr}(z)$ (purple dashed, right axis), for a vertically isothermal thin disk as describe in section \ref{vertdisk}. The x-axis is $z/H$. Curves show three imposed radial gradients of the mean vertical field, $\partial_r B_z=\{0.01,\,0.1,\,0.3\}$. $D_\phi$ implements equation(\ref{eq:diffusion}) with stresses $R_{ij}\propto \alpha P$ and $M_{ij}/\rho\propto \alpha c_s^{2}$, while $\alpha_{rr}$ follows from the helicity constraint, equation(\ref{eq:Alpha_ij}). We run the adapted parameters $\alpha=0.1$, $\tau_c=1$, $\bar Q_{rr}=0.7$, $\bar Q_{zz}=0.9$, $\bar Q_{\phi\phi}=0.1$, $\bar Q_{r\phi}=0.05$, and a toroidal field envelope $B_\phi/B_0=0.2$. Increasing $\partial_r B_z$ boosts $|D_\phi|$ near the midplane and drives larger $|\alpha_{rr}|$, illustrating the locking between transport and helicity predicted by the model as expected.}
  \label{fig:Dphi_alpha}
\end{figure*}

Figure~\ref{fig:Dphi_alpha} delineates the vertical transport–dynamo coupling predicted by our closure as describe in section \ref{vertdisk}. For three imposed radial gradients of the vertical field, $\partial_r B_z=\{0.01,\,0.1,\,0.3\}$, the azimuthal turbulent diffusion $D_\phi(z)$, (see equation(\ref{eq:Diff_Azm})) peaks within $|z|\!\lesssim\!H$ and grows systematically with $\partial_r B_z$; for the larger gradients it exhibits sign reversals across $z/H$, indicating layers of effective anti–diffusion and convergent radial transport of $B_z$. The helicity–regulated coefficient $\alpha_{rr}(z)$ (refer equation \ref{eq:alphaij_diff}) co–varies with these changes and adjusts to maintain the helicity–transfer balance $B_i\alpha_{ij}B_j \simeq -\,B_q D_{qsu}\partial_s B_u$, linking dynamo backreaction directly to flux transport. The combined trends demonstrate that modest changes in the large–scale radial field gradient reorganize both the vertical magnetic geometry and the effective transport, with the strongest departures from scalar–diffusion behavior concentrated near the midplane where MRI turbulence is most anisotropic.

\subsection{Interpretation of the Turbulent Transport Equations}

From the section above, in particular, equation(\ref{eq:Diff_Azm}), provides an expression for the azimuthal component of the turbulent electromotive force, incorporating gradients of the magnetic field components and tensorial coefficients representing the turbulent Reynolds and Maxwell stresses, as well as eddy anisotropy via $\bar{Q}_{ij}$. Each term encodes, vertical gradients $\partial_z B_r$ and $\partial_z B_\phi$ dominate in the disk corona, while radial gradients $\partial_r B_z$ become significant in regions of substantial midplane accretion \citep[e.g.,][]{JV18}. 

Crucially, the term involving $(1 - 2\bar{Q}_{ij})$ modulates the effective diffusivity. When the anisotropy tensor $\bar{Q}_{ij}$ becomes large in a particular direction for azimuthally elongated eddies with $\bar{Q}_{\phi\phi} \ll \bar{Q}_{zz}, \bar{Q}_{rr}$—the effective diffusivity in that direction can become negative. Thus, this has a significant influence on the result; rather than dissipating magnetic structures, the turbulence acts to concentrate magnetic flux, a process often referred to as \emph{anti-diffusion}. Such behavior is commonly observed in the vicinity of the disk midplane, where shallow magnetic field gradients align with pronounced turbulent anisotropy \citep{Brandenburg_Subramanian05,Lesur_Ogilvie08,Vishniac_Shapovalov14,SquireJ+25}. The helicity-driven $\alpha_{ij}$ term in equation \ref{eq:Alpha_ij}, which is a feedback mechanism resulting from the near-conservation of magnetic helicity in high magnetic Reynolds number systems. Magnetic helicity, a topological measure of the knottedness and linkage of field lines, cannot cascade to small scales as efficiently as energy. As a result, large-scale field evolution becomes constrained by helicity conservation. In this context, turbulent eddies accumulate helicity when the large-scale $\alpha$ effect is active, eventually producing a back-reaction that quenches further field amplification \citep{Lesur_Longaretti11,Vishniac_Cho01,Vishniac_Shapovalov14,Rodman&Reynolds24}.

This feedback is mathematically expressed in equation(\ref{eq:alphaij_diff}), where the turbulent diffusivity term $D_{ijk}$ and the induced $\alpha_{ij}$ are locked together through the requirement that the net helicity transfer vanishes in the stationary state. Here, helicity conservation provides a nonlinear integral constraint, rather than a mere scaling relation \citep{Vishniac_Shapovalov14,Zenati_Vishniac23}. Taken together, these equations suggest that the long-term structure of magnetic fields in accretion disks can not be understood without including anisotropic turbulence and helicity regulation. The model thus provides a pathway toward resolving longstanding puzzles in disk theory, such as the origin of large-scale vertical magnetic fields, the saturation of the MRI, and the emergence of magnetically dominated corona.

\begin{figure*}
    \centering
   \includegraphics[width=1\linewidth]{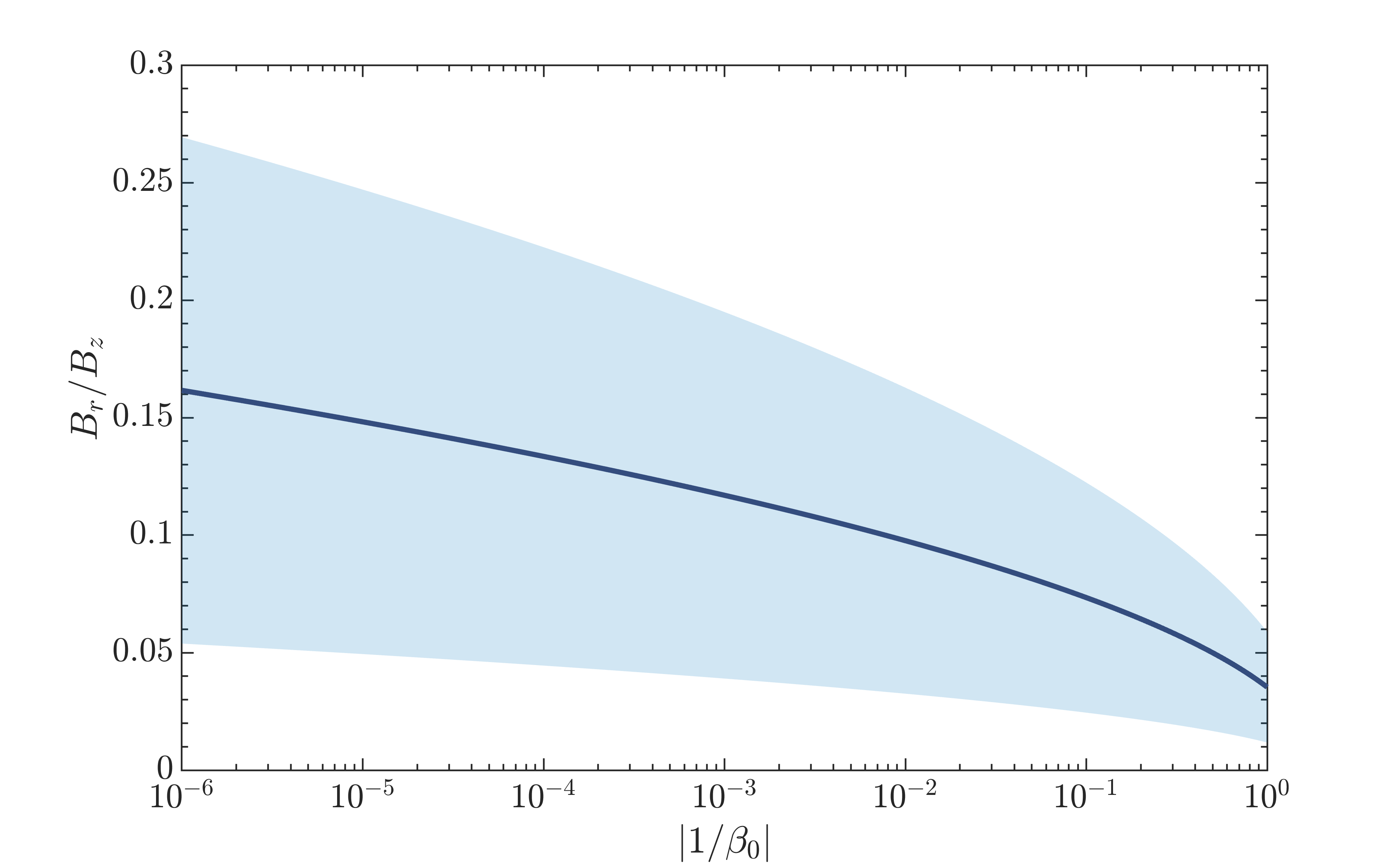}
    \caption{Radial bending of the large-scale magnetic field at the thin disk surface. The $B_r/B_z$ as a function of the magnetization parameter $\beta_0^{-1} = B_z^2/(4\pi P_0)$ adapted the density profile as in Figure \ref{fig:Dphi_alpha} with different $H/R$ initial values.}
  \label{fig:BrBz_beta}
\end{figure*}

\section{Relevance to Observations} \label{sec:obesr}

A key challenge in disk theory is reconciling the presence of sustained vertical fields required to launch magnetocentrifugal winds and jets with the destructive tendencies of turbulence and reconnection \citep{Alexander&Axel11,SalvesenG+2016,LiskaM+2020}. Our model provides a mechanism for stabilizing and even enhancing vertical fields through anisotropic anti-diffusion and helicity regulation.

In protostellar disks, observations from ALMA have revealed ringed substructures and polarized emission indicative of coherent magnetic field morphology. These features are difficult to reproduce in models with purely turbulent diffusion. The formalism presented here explains how field lines can remain organized across large radial scales, particularly in regions where the MRI is active but anisotropic. The anti-diffusive behavior near the midplane, for instance, can lead to the formation of a magnetic ring that mirrors observed dust rings.

In low-mass X-ray binaries (LMXBs) and active galactic nuclei (AGNs), the episodic emergence of jets and coronae points to intermittent field amplification and vertical flux accumulation. The helicity-regulated $\alpha$ term derived in this work provides a natural explanation for this behavior. As turbulent eddies accumulate helicity, the $\alpha$ effect becomes suppressed, halting further flux growth and leading to quasi-cyclic activity. This can help explain the spectral state transitions in black hole binaries and AGNs, where changes in magnetic configuration appear to play a key role. Moreover, recent polarimetric studies of circumplanetary and circumbinary disks show large-scale alignment of magnetic field vectors, often with a vertical or poloidal bias. These morphologies support theoretical models suggesting that vertical magnetic flux is not fully diminished by turbulence and reconnection, but is instead preserved and redistributed through anisotropic turbulence, magnetic buoyancy, and helicity-conserving feedback.

The presence of external field lines suggests the possibility of significant angular momentum loss through the embedded poloidal field. Comparing the MRI and large scale field terms in equation(\ref{eq:amflux}) we see that the former has former is suppressed by a factor of $\sim H/r$ since the latter has an effective surface for emission of $r^2$ whereas turbulent transport in the disk operates across a surface are of $Hr$. The ability of the disk to sustain bending angles of order unity means that angular momentum loss through the magnetosphere can dominate over internal transport even while the magnetic pressure near the disk midplane can be of order $H/r$ times the gas pressure. We have not tried to construct global thin disk models in this paper, but this suggests that $\alpha$ models of the inner parts of accretion disks are likely to overestimate the accretion time and overestimate the gas column density. This may be particularly important for modeling dwarf novae outbursts and other systems that cycle through states due to a lack of a stable accretion solution corresponding to their average mass accretion rate \cite[see,][]{ScepiN+19}.

Finally, the predicted variation of the turbulent transport coefficients with disk height and radial location implies that the field evolution will be spatially heterogeneous—consistent with multiwavelength variability and patchy jet launching regions observed in both stellar and galactic systems.

\section{Conclusion} \label{sec:conclusion}

In this research, we provide a robust theoretical foundation for interpreting magnetic phenomena observed in protostellar systems, X-ray binaries, and AGNs. The model predicts spatially non-uniform magnetic diffusion and the formation of self-organizing flux concentrations, which align with observational characteristics such as magnetic rings, jet intermittency, and polarimetric asymmetries. This alignment underscores the model's relevance and enhances its applicability to complex astrophysical processes. 

We have examined the evolution of magnetic fields in thin, turbulent accretion disks by analyzing a set of coupled equations for turbulent diffusion and helicity-regulated field transport. Equations \ref{eq:Diff_Azm} - \ref{eq:alphaij_diff} provide a generalized formulation of the turbulent electromotive force, incorporating both anisotropic diffusion and dynamical $\alpha$-effects arising from magnetic helicity conservation. This phenomenon provides a compelling explanation for the persistence and organization of large-scale poloidal fields despite the presence of significant turbulence.

The vertical profiles of $D_{ijk}$ and $\alpha_{rr}$ highlight the sensitivity of magnetic transport to local gradients in $B_z$ and reveal how moderate radial field variations can dramatically reshape the field topology. These results provide a quantitative explanation for the emergence of vertically stratified magnetic structures, midplane flux accumulation, and field saturation observed in numerical simulations and inferred from disk observations.

In summary, we emphasize that the thin-disk constraint $E_\phi=0$ does not, by itself, select a unique steady distribution of vertical flux. In a kinematic thin-disk treatment, the exact condition can be satisfied by a continuous family of balances among inward advection of $B_z$, outward radial turbulent diffusion, and vertical transport of the inclined field, including solutions with $B_r = B_\phi = 0$ if $B_z(r)$ is tuned so that advection and diffusion cancel. Therefore, the actual poloidal-field bending cannot be determined solely from the induction constraint; it requires physics that couples the disk interior to the low-density surface layers and to the external field. In practice, this coupling is supplied by magnetic tension; any surface inclination must communicate inward as a systematic, tension-driven drift that is negligible near the midplane but becomes dominant toward the surface where gas support weakens. Including this effect removes the degeneracy, rendering the bending angle a dynamically selected outcome rather than a free parameter.

Furthermore, the incorporation of a helicity constraint introduces a nonlinear feedback mechanism that regulates the dynamo process. This feedback effectively saturates the $\alpha-$effect, thereby preventing the unbounded growth of magnetic energy. The regulation, as articulated in equation(\ref{eq:Alpha_ij}), establishes a connection between the evolution of large-scale magnetic fields and the dynamics of small-scale helicity in a coherent and physically significant manner. The generalized constraint equation(\ref{eq:alphaij_diff}) indicates that in a stationary state, the turbulent transfer of helicity diminishes due to the interplay between the $\alpha-$effect and the anisotropic diffusion term. We confidently assert that the regulation of helicity is crucial in preventing the unrestrained growth of magnetic energy, thereby ensuring the stability of large-scale poloidal field configurations.

The presence of external field lines suggests the possibility of significant angular momentum loss through the embedded poloidal field. Comparing the MRI and large scale field terms in equation(\ref{eq:Induc}).

Finally, this study synthesizes previous research on turbulent transport and helicity regulation, providing a robust framework to promote stable magnetic field configurations in astrophysical thin disks.

\section*{Acknowledgments}

Y.Z. acknowledges the visitor support of the Kavli Institute for Cosmology at Cambridge, where part of this work was completed. ETV acknowledges the support of the AAS journals.\\
%acknowledge visitor support from the Kavli Institute for Cosmology, Cambridge, where part of this work was completed.\\

\bibliography{RefMHD2024}
\end{document}